\documentclass[a4paper,11pt]{article}
\pdfoutput=1 

\usepackage{jheppub} 

\usepackage[T1]{fontenc} 
\usepackage{multirow}
\usepackage{graphicx}
\usepackage{appendix}

\title{
    \vspace{-4cm}\hfill {\small  IPPP/22/76 \\ \vspace{-0.3cm} \hfill }\vspace{3cm}\\ 
   Multi-variable Integration with a Neural Network}

\author[a,1]{D. Ma\^{\i}tre, \note{Corresponding author.}}
\author[b]{R. Santos-Mateos}

\affiliation[a]{Institute for Particle Physics Phenomenology, Durham University, Durham DH1 3LE, UK}
\affiliation[b]{Department of Electronics and Computing, University of Santiago de Compostela, Spain}

\emailAdd{daniel.maitre@durham.ac.uk}

\abstract{
   In this article we present a method for automatic integration of parametric integrals over the unit hypercube using a neural network. The method fits a neural network to the primitive of the integrand using a loss function designed to minimize the difference between multiple derivatives of the network and the function to be integrated. We apply this method to two example integrals resulting from the sector decomposition of a one-loop and two-loop scalar integrals. Our method can achieve per-mil and percent accuracy for these integrals over a range of invariant values. Once the neural network is fitted, the evaluation of the integral is between 40 and 125 times faster than the usual numerical integration method for our examples, and we expect the speed gain to increase with the complexity of the integrand. 
}

\begin{document}
\maketitle
\flushbottom
\section{Introduction}
The problem we consider is that of efficiently computing $k$-dimensional parametric integrals of the type 
\begin{equation}\label{eq:integral}
I(s_1,...,s_m) = \int\limits_0^1dx_1 \dots \int\limits_0^1dx_k \;f(s_1,...,s_m; x_1,...,x_k)\;.
\end{equation}
where the variables $x_i$ are auxiliary variables to be integrated over and $s_i$ are parameters that are not integrated over. 

\begin{figure}
    \centering
    \includegraphics[scale=0.46]{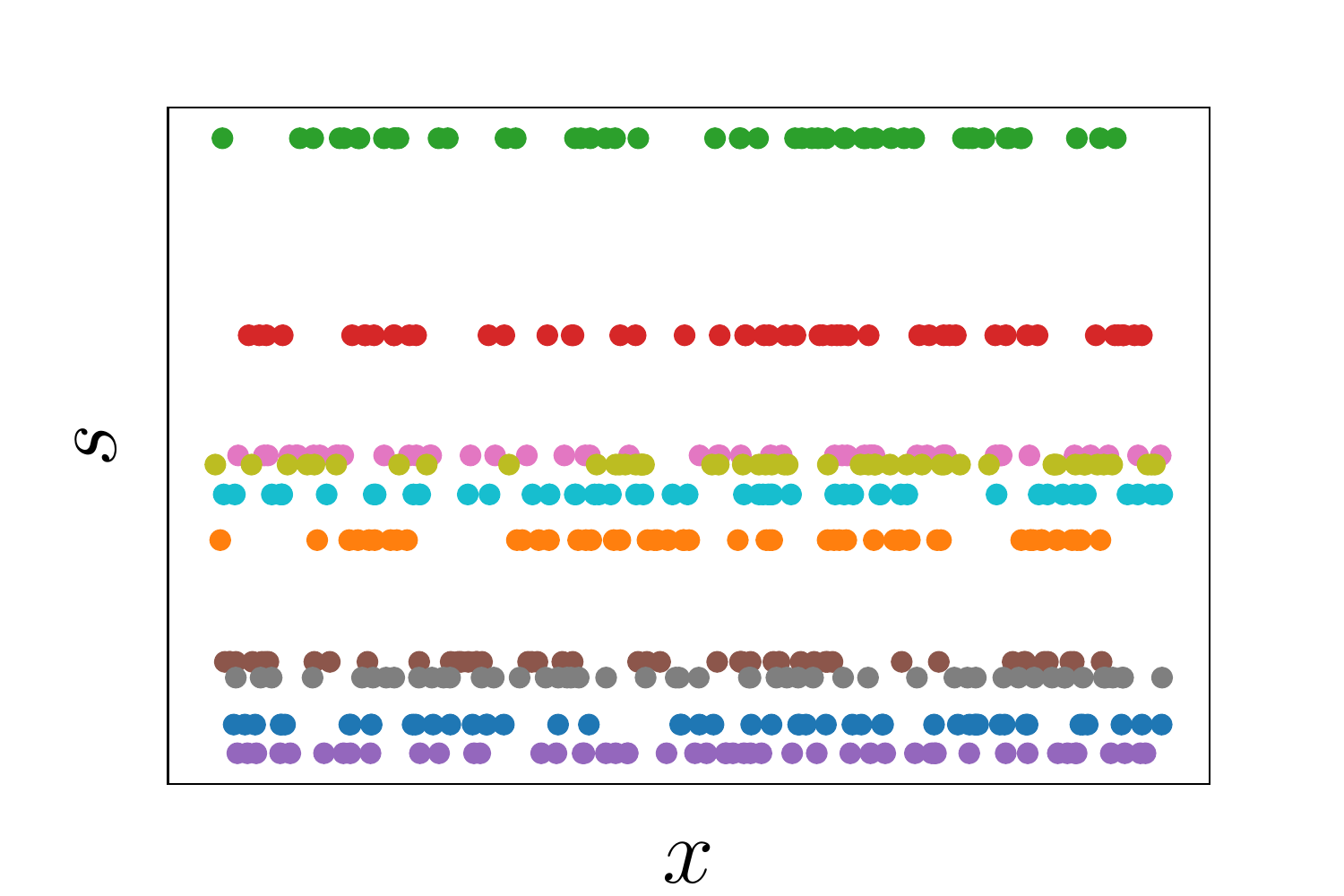}
    \caption{Illustration of the sampling in $s$-$x$ space. For independent numerical integrations each line of different color corresponds to an individual integration and no information is pooled between them. }
    \label{fig:illustration1}
\end{figure}

For typical integrands $f$ the integration cannot be performed analytically and a numerical integration is required, using Monte Carlo or Quasi Monte Carlo methods. If the value of the function $I$ is required for a large number of different values of $s_1,...,s_m$ many such numerical integrations have to be performed. These MC integrations are performed independently and any information about the integrand gathered for one integration for a given set of parameters $s_1,...,s_m$ is not leveraged for the integration for new parameters $s'_1,...,s'_m$, even if the new parameters only change the integrand values in a very mild manner. This is illustrated in figure~\ref{fig:illustration1}: each evaluation of the function $I$ requires a large number of evaluations with a range of values of $x$ and with these chosen values of $s$. 

\begin{figure}
    \centering
    \includegraphics[scale=0.46]{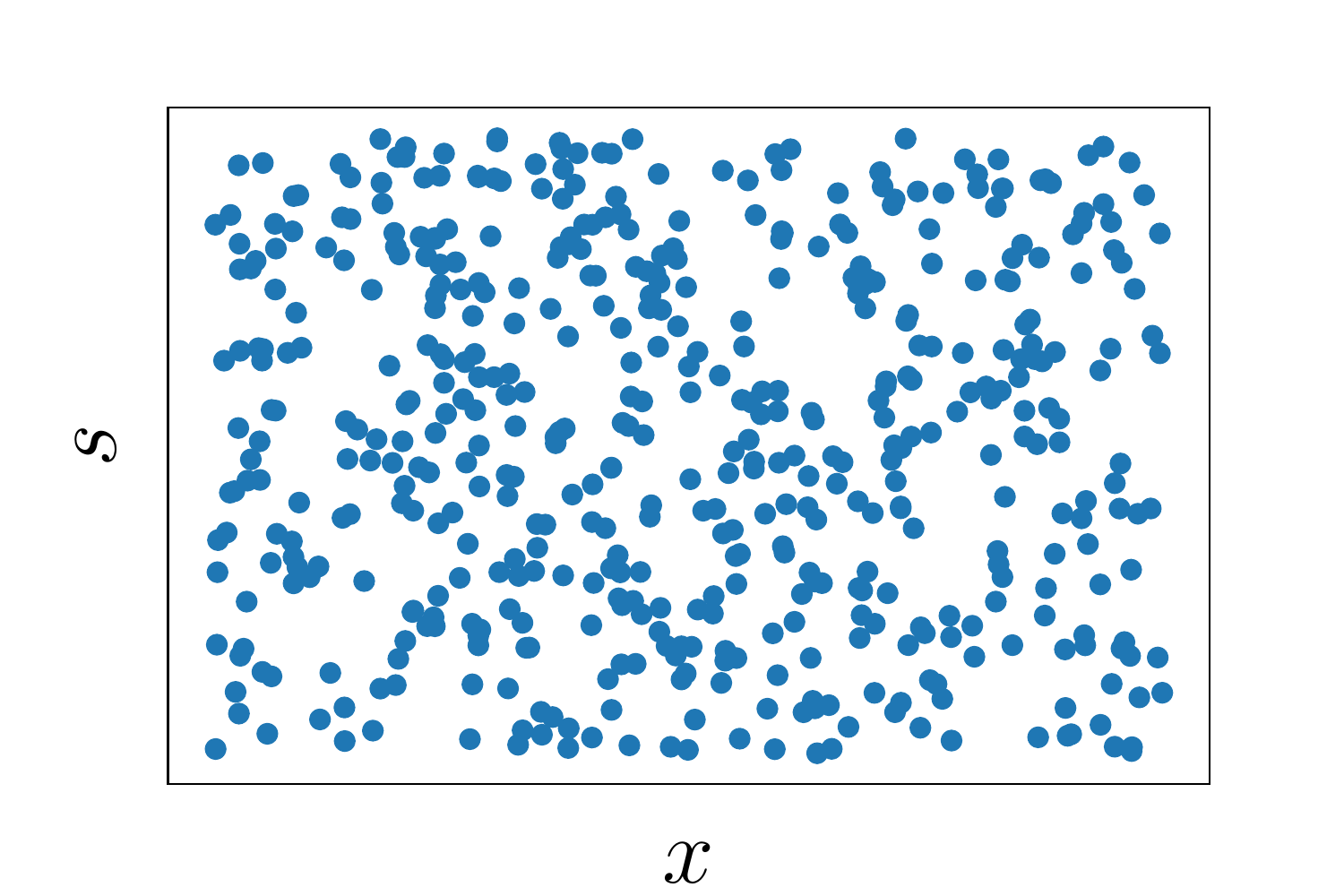}
    \includegraphics[scale=0.46]{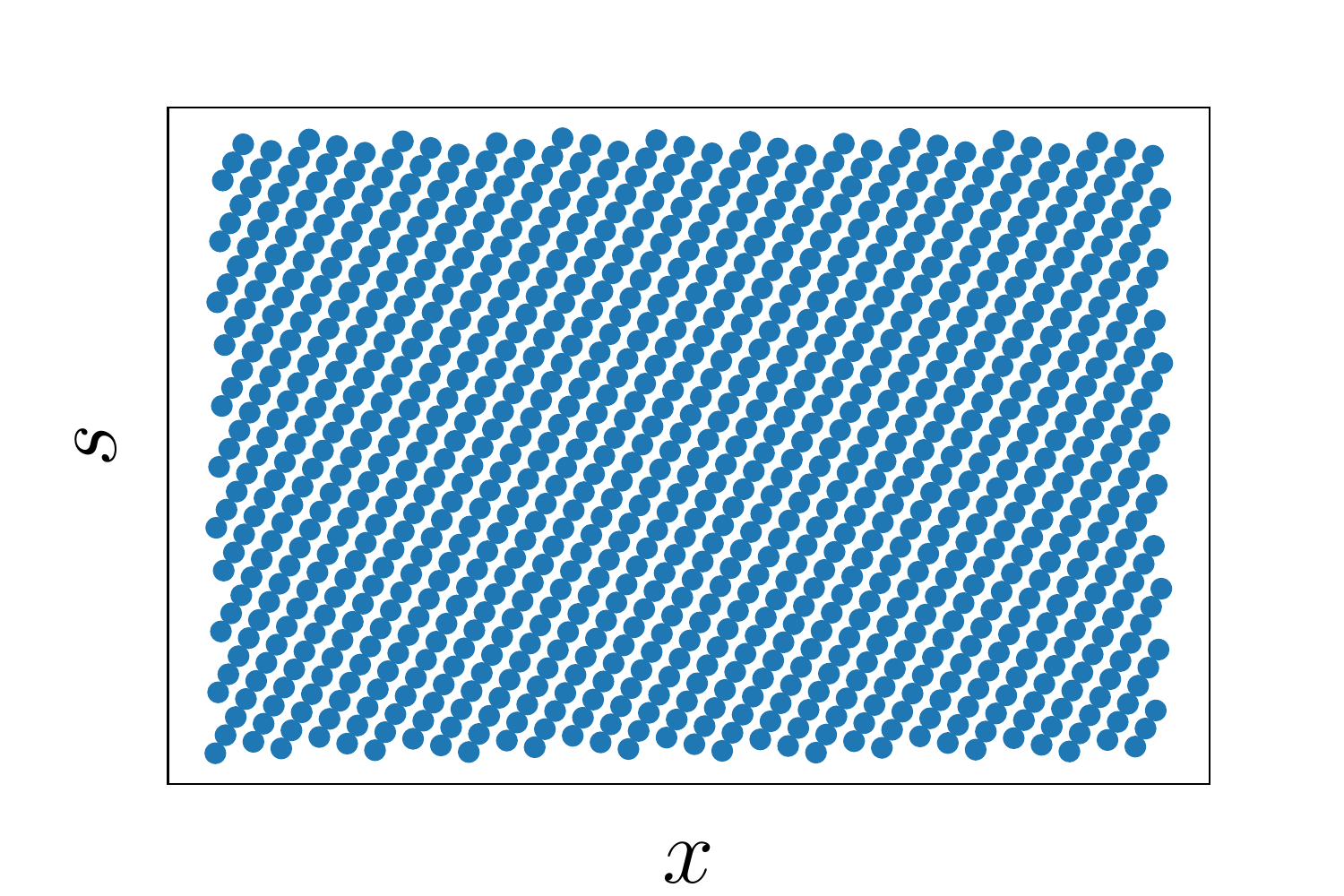}
    \caption{Illustration of the point selection in $s$-$x$ space for random (left) and Quasi-Monte Carlo (right) sampling.}
    \label{fig:illustration2}
\end{figure}

In this work we aim at sampling the $s$-$x$ space uniformly in order to use information from the smoothness of $f$ (and therefore $I$) as a function of the $s$ variables. In essence we will find an estimator that mimics the integrand $f$ in $s$-$x$ space, as illustrated in Fig.~\ref{fig:illustration2}, but additionally we will use a form for this estimator that allows us to calculate the integration over the $x$ variables analytically, in an exact and efficient fashion.


Using neural networks to produce integrated quantities has been applied in other fields. In material science it has been used to estimate the free energy density from single differential data~\cite{TEICHERT2019201}, or in image rendering~\cite{DBLP:journals/corr/abs-2012-01714}. These differ from our work in that the former uses information on individual partial derivatives with respect  to multiple variables, and in the latter it is applied to a single variable integration\footnote{The authors mention the extension to multiple integrations in the supplemental material.}. In Ref.~\cite{9084130} the authors use a neural network to automatically integrate a function over several variables. The difference with our approach is that the representability of the integrand is limited by the single-layer nature of their network and the fact that they considered the integration over all argument of the function as opposed to our setup where we consider parametric integrals. In contrast we use a deeper network to increase the representability for the integrand, and incorporate parametric dependence of the integrand, at the cost of implementing a more complicated loss function.    

In the field of particle physics neural networks and other ML techniques have been used to improve the efficiency of Monte Carlo integrations~\cite{Bendavid:2017zhk,Gao:2020vdv,Klimek:2018mza,Bothmann:2020ywa,Verheyen:2020bjw,Chen_2021} and event generation~\cite{Gao:2020zvv,Otten:2019hhl,Hashemi:2019fkn,DiSipio:2019imz,Butter:2019cae,Bishara:2019iwh,Backes:2020vka,Butter:2020qhk,Alanazi:2020klf,Nachman:2020fff}. In this work we attempt something different: instead of improving the Monte Carlo integration we aim at \emph{replacing} it with a fitting process. 

The paper is structured as follows: Section~\ref{sec:method} outlines the method we used and derives the required quantities. Section~\ref{sec:results} shows our result for two test cases arising from the sector decomposition of one- and two-loop scalar integrals for the gluon scattering into two Higgs bosons. 
We conclude in Section~\ref{sec:conclusion}.

\section{Method overview}\label{sec:method}
If the integrand $f$ was simple enough to integrate analytically, we could compute the primitive $F$ of $f$ such that 
\begin{equation}
    \frac{d^{k}F(s_1,...,s_m;x_1,...,x_k)}{dx_1\dots dx_k} = f(s_1,...,s_m;x_1,...,x_k)
\end{equation}

and we could obtain the value of the integral $I$ in Equ.~\ref{eq:integral} by integrating and evaluating at the integration boundaries 
\begin{eqnarray}
    \lefteqn{I(s_1,...,s_m)=}\nonumber\\
    &=&  \int\limits_0^1dx_1 \dots \int\limits_0^1dx_k \;f(s; x_1,...,x_k)
=\int\limits_0^1dx_1 \dots \int\limits_0^1dx_k \;\frac{d^{k}F(s; x_1,...,x_k)}{dx_1\dots dx_k}\nonumber\\
&=&  \int\limits_0^1dx_1 \dots \int\limits_0^1dx_{k-1} \;\frac{d^{k-1}F(s; x_1,...,x_{k-1},1)}{dx_1\dots dx_{k-1}} - \frac{d^{k-1}F(s; x_1,...,x_{k-1},0)}{dx_1\dots dx_{k-1}}\;,\nonumber \\
\end{eqnarray}
where we have used the shorthand $s\equiv s_1,...,s_m$. Repeating the above steps $k-1$ times we find
\begin{equation}\label{eq:boundarysum}
    I(s_1,...,s_m) = \sum\limits_{x_1,...,x_k=0,1} (-1)^{k-\sum x_i}F(s_1,...,s_m; x_1,...,x_k)\,.
\end{equation}

In any application of interest $f$ can be evaluated easily but a primitive $F$ cannot be calculated analytically. 

The idea we present in this paper is to use a neural network 

\begin{equation}
\mathcal{N}(s_1,...,s_m;x_1,...,x_k)    
\end{equation}
to provide an approximation for the primitive $F$. It is important to note that the auxiliary variables $x_i$ and the parameters $s_i$ are treated as equal inputs to the network, the difference will be in how they are treated in the loss function. The network will be trained such that its $k$-th derivative with respect to the auxilliary variables $x_1,...x_k$ matches the integrand $f$, using a loss function $L$ given by
\begin{equation}\label{eq:loss}
L={\rm MSE}\left(f(s_1,...,s_m;x_1,...,x_k), \frac{d\mathcal{N}(s_1,...,s_m;x_1,...,x_k)}{dx_1...dx_k}\right)\,.
\end{equation}
To construct a network whose output can be differentiated multiple times with respect to some of its inputs we consider a family of function $f_n(x)$ such that 

\begin{equation}
    \frac{df_n(x)}{dx}=f_{n-1}(x)
\end{equation}
and require all functions $f_n$ to be continuous for $0\leq n\leq k$. There is some freedom in the choice of functions for this family. One can choose $f_0$ to be a regular activation function. If $f_0$ is the sigmoid function then
\begin{equation}
    f_0(x) = \frac{1}{1+e^{-x}}\;,\quad f_1(x) = \log(1+e^x)\;,\quad f_n(x) = -\rm{Li}_n(-e^x)
\end{equation}
and if we choose $f_0$ to be the ReLU activation function we have
\begin{equation}
    f_n(x)=0 \quad \mbox{if}\quad  x<0\;,\qquad f_n(x) = \frac{1}{n!}x^n\quad\mbox{if}\quad x\geq 0\;.
\end{equation}
Alternatively one can pick $f_k$ to be a regular activation function, such as $\tanh$ or the sigmoid function. For the results presented in this paper we use the sigmoid function as our function $f_k$. Appendix~\ref{sec:sigmoidderivatives} gives the derivatives of the sigmoid and $tanh$ functions needed to build the family $f_n$. 

To construct the neural network approximation $\mathcal{N}$ of the primitive of $F$ we use a fully connected network with activation function $f_k$, such that after differentiating $k$ times we obtain combinations of the members of the family of functions $f_n$. \footnote{It is important to start with activation function in $\mathcal{N}$ that can be differentiated multiple times. Imagine we were to use ReLu in $\mathcal{N}$, upon differentiation we would obtain vanishing activations for the derivatives (except for a delta function at zero argument that would not be implementable) and we would be unable to fit it to the derivative data.}

\subsection{Calculating the derivatives}

In this section we derive the expressions for the derivative of the neural network that we need in the loss function Eq.~(\ref{eq:loss}).

We denote the output of node $i$ of layer $l$ with $a_i^{(l)}$, it is given by
\begin{equation}
    a^{(l)}_i = \phi\left(z^{(l)}_i\right)\;,\quad z^{(l)}_i = \sum_j w^{(l)}_{ij} a_j^{(l-1)} +b_i^{l} \;.
\end{equation}
where $\phi$ is the activation function. The first layer is a special case with
\begin{equation}
a^{(0)}_i=x_i\;\mbox{for}\;i\leq k\;,\quad a^{(0)}_i=s_{i-k}\;\mbox{for}\;i > k\;.  
\end{equation}

The output of the network with $L$ layers is given by
\begin{equation}
    y = \sum_j w^{(L+1)}_{j} a_j^{(L)}+ b^{(L)}
\end{equation}

The ingredients we need for the derivative of the output with respect to the first input $x_1$ are
\begin{equation}
    \frac{dy}{dx_1} = \sum\limits_j w^{(L+1)}_j\frac{d a_j^{(L)}}{dx_1}
\end{equation} 
where the derivative of the activation at level $l$ is given by
\begin{equation}
    \frac{da^{(l)}_i}{dx_1} = \phi'(z^{(l)}_i) \frac{dz_i^{(l)}}{dx_1}= \phi'(z^{(l)}_i)\left(\sum\limits_j w^{(l)}_{ij}\frac{d a_j^{(l-1)}}{dx_1}\right)\;,
\end{equation} 
with the special case for $l=0$: 
\begin{equation}
    \frac{da^{(0)}_i}{dx_1} = \delta_{1i}
\end{equation}

Now we differentiate again, this time with respect to the second input $x_2$:

\begin{equation}
    \frac{d^2y}{dx_1dx_2} = \sum\limits_j w^{(L+1)}_j\frac{d^2 a_j^{(L)}}{dx_1dx_2}\;.
\end{equation} 
We need the double differential of the activations for each layer:
\begin{eqnarray}
    \frac{d^2a^{(l)}_i}{dx_1dx_2} &=& \phi''(z^{(l)}_i) \frac{dz_i^{(l)}}{dx_1}\frac{dz_i^{(l)}}{dx_2} + \phi'(z^{(l)}_i) \frac{d^2z_i^{(l)}}{dx_1dx_2}= \nonumber\\
    &=& \phi''(z^{(l)}_i)\left(\sum\limits_j w^{(l)}_{ij}\frac{da_j^{(l-1)}}{dx_1}\right)\left(\sum\limits_j w^{(l)}_{ij}\frac{d a_j^{(l-1)}}{d x_2}\right) \nonumber\\
    &&+\phi'(z^{(l)}_i)\left(\sum\limits_j w^{(l)}_{ij}\frac{d^2 a_j^{(l-1)}}{dx_1 dx_2}\right)
\end{eqnarray} 
with the special case for the input layer: 
\begin{equation}
    \frac{d^2a^{(0)}_i}{dx_1dx_2} = 0\;.
\end{equation}
We observe that differentiating twice yields an expression that mixes the differential of the activation function to different order. Repeating the process $k$ time gives the expression we need for the loss function in Equation~\ref{eq:loss}. Appendix~\ref{sec:derivatives} shows explicit expressions for the third and fourth derivatives. 

When starting with a neural network $\mathcal{N}$ with activation function $f_k$ we have 
\begin{equation}
    \phi^{(i)} = f_{k-i}\;.
\end{equation}

\subsection{Preprocessing}
The values of the integrand can span a wide range for different values of the parameters $s_1,...,s_m$, making the fitting more difficult.\footnote{For example multiplying all parameters $s_{12}, s_{14}, m_t^2, m_H^2$ by a factor of 2 reduces the value of the integrand $I_2$ by a factor of 8.} To alleviate this problem we found it useful to normalise the integrand by its value at a fixed location in $x$ space, which we chose to be the centre of the unit hypercube. In practice this means fitting to a modified integrand
\begin{equation}
    f\rightarrow \tilde f(s_1,...,s_m;x_1,...,x_k)\equiv \frac{f(s_1,...,s_m;x_1,...,x_k)}{f(s_1,...,s_m; \frac{1}{2},\frac{1}{2},,...,\frac{1}{2})} 
\end{equation}
and the estimate of the integral will be normalised 
\begin{equation}
    I\rightarrow \tilde I(s_1,...,s_m)\equiv \frac{I(s_1,...,s_m)}{f(s_1,...,s_m; \frac{1}{2},\frac{1}{2},...,\frac{1}{2})} \,.
\end{equation}

We found that applying a Korborov transformation~\cite{korobov} improves the accuracy of the method. The transform is defined through a weight function $w$ normalised such that 
\begin{equation}
    \int\limits_0^1 w(t) dt = 1\;.
\end{equation}
With this weight function we can define the variable transform 
\begin{equation}
    x(t) = \int\limits_0^t w(t') dt'
\end{equation}
and inserting in the integral definition we get
\begin{equation}
    \int\limits_0^1 dx f(x) = \int\limits_0^1 dt\, w(t) f(x(t))    \;. 
\end{equation}
For the work presented in this article we use the weight function and transform
\begin{equation}
w(t) = 6t(1-t)\;, \quad x = t^2  (3-2 t) 
\end{equation}
for each of the $x_i$ variables. This choice makes the integrand vanish at the $x_i$ boundaries.

\subsection{Training}
The loss in equation~(\ref{eq:loss}) can be constructed using the derivatives in the above section. The parameters $w_{ij}^{(l)}$ can be learned using gradient descent or any other optimizer, for this paper we used the Adam optimizer~\cite{adam}. In our implementation we use the autodiff feature of PyTorch~\cite{NEURIPS2019_9015}, but for a more efficient implementation one could obtain explicit formulae for the gradients by differentiating the loss with respect to the network parameters $w^{(l)}_{ij}$, resulting in an analog of back-propagation, but also involving derivatives of the activation function. 

Since we are fitting the network derivatives to exact values of the integrand, there is no noise in the data and no need for regularisation to prevent fitting noisy behaviour in the data. Since in our case the integrand $f$ is comparatively cheap to calculate we can ensure a good level of generalisation by training the network with a very large number of different values of $x_1,...,x_k$ and $s_1,...,s_m$. The luxury of being able to generate as large a training set as required is setting this application apart from much of the typical machine learning literature. Another aspect of this application that differs from more common practice is that we pushing the precision of the network much further than in conventional uses of neural networks for regression tasks. 

\section{Results}\label{sec:results}
To showcase our method we apply it to the parametric integral
\begin{equation}\label{eq:1lbox}
I_1(s_{12}, s_{14}, m_H^2, m_t^2) = \int\limits_0^1 dx_1\int\limits_0^1 dx_2\int\limits_0^1 dx_3 \frac{1}{F_1^2}
\end{equation}
with 
\begin{eqnarray}
    F_1 &=& m_t^2+2 x_3 m_t^2+x_3^2 m_t^2+2 x_2 m_t^2-x_2 s_{14}+2 x_2 x_3 m_t^2\nonumber\\
    &&-x_2 x_3 m_H^2+x_2^2 m_t^2
    +2 x_1 m_t^2+2 x_1 x_3 m_t^2 \nonumber\\
    &&- x_1 x_3 s_{12}+2 x_1 x_2 m_t^2-x_1 x_2 m_H^2 +x_1^2 m_t^2\;,
\end{eqnarray}
which arises after sector decomposition of a scalar integral for the one-loop box for the $gg\rightarrow hh$ process. 

The second integrand arises from the sector decomposition from a two-loop box integral for the same process:
\begin{equation}
I_2( s_{12}, s_{14},m_H^2,m_t^2) =
 \int\limits_0^1 dx_1\int\limits_0^1 dx_2\int\limits_0^1 dx_3 \int\limits_0^1 dx_4\int\limits_0^1 dx_5\int\limits_0^1 dx_6\frac{2U_2 x_4}{F_2^3}\,,
\end{equation}
with\begin{eqnarray}
\lefteqn{F_2=}&&\nonumber\\
&&m_t^2+x_6 m_t^2+2 x_5 m_t^2-x_5  s_{12}+2 x_5 x_6 m_t^2-x_5 x_6  s_{12}+x_5^2 m_t^2+x_5^2 x_6 m_t^2+x_4 m_t^2+2 x_4 x_6 m_t^2\nonumber\\
&&-x_4 x_6  s_{12}+x_4 x_5 m_t^2+2 x_4 x_5 x_6 m_t^2 +x_4^2 x_6 m_t^2+x_3 m_t^2+2 x_3 x_5 m_t^2-x_3 x_5  s_{12}+x_3 x_5^2 m_t^2\nonumber\\
&&+2 x_3 x_4 m_t^2-x_3 x_4 m_H^2+2 x_3 x_4 x_6 m_t^2-x_3 x_4 x_6 m_H^2+2 x_3 x_4 x_5 m_t^2-x_3 x_4 x_5 m_H^2\nonumber\\
&&+2 x_3 x_4 x_5 x_6 m_t^2-x_3 x_4 x_5 x_6 m_H^2+2 x_3 x_4^2 x_6 m_t^2-x_3 x_4^2 x_6 m_H^2+x_3^2 x_4 m_t^2+x_3^2 x_4 x_5 m_t^2\nonumber\\
&&+x_3^2 x_4^2 x_6 m_t^2+x_2 m_t^2+2 x_2 x_5 m_t^2-x_2 x_5  s_{12}+x_2 x_5^2 m_t^2+2 x_2 x_4 m_t^2-x_2 x_4  s_{12}+2 x_2 x_4 x_6 m_t^2\nonumber\\
&&+2 x_2 x_4 x_5 m_t^2-x_2 x_4 x_5  s_{12}+2 x_2 x_4 x_5 x_6 m_t^2-x_2 x_4 x_5 x_6  s_{12}+2 x_2 x_4^2 x_6 m_t^2\nonumber\\
&&-x_2 x_4^2 x_6  s_{12}+2 x_2 x_3 x_4 m_t^2-x_2 x_3 x_4 m_H^2+2 x_2 x_3 x_4 x_5 m_t^2-x_2 x_3 x_4 x_5 m_H^2\nonumber\\
&&+2 x_2 x_3 x_4^2 x_6 m_t^2-x_2 x_3 x_4^2 x_6 m_H^2+x_2^2 x_4 m_t^2+x_2^2 x_4 x_5 m_t^2+x_2^2 x_4^2 x_6 m_t^2+2 x_1 m_t^2\nonumber\\
&&+2 x_1 x_6 m_t^2 +2 x_1 x_5 m_t^2+2 x_1 x_5 x_6 m_t^2+x_1 x_4 m_t^2+2 x_1 x_4 x_6 m_t^2+2 x_1 x_3 m_t^2+2 x_1 x_3 x_5 m_t^2\nonumber\\
&&+2 x_1 x_3 x_4 m_t^2-x_1 x_3 x_4 m_H^2+2 x_1 x_3 x_4 x_6 m_t^2-x_1 x_3 x_4 x_6  s_{14}+x_1 x_3^2 x_4 m_t^2+2 x_1 x_2 m_t^2\nonumber\\
&&+2 x_1 x_2 x_5 m_t^2+2 x_1 x_2 x_4 m_t^2-x_1 x_2 x_4  s_{12}+2 x_1 x_2 x_4 x_6 m_t^2+2 x_1 x_2 x_3 x_4 m_t^2\nonumber\\
&&-x_1 x_2 x_3 x_4 m_H^2+x_1 x_2^2 x_4 m_t^2+x_1^2 m_t^2+x_1^2 x_6 m_t^2+x_1^2 x_3 m_t^2+x_1^2 x_2 m_t^2
    \end{eqnarray}
and 
\begin{eqnarray}
    U_2&=& 1+x_6+x_5+x_5 x_6+ x_4 x_6+ x_3+ x_3 x_5+ x_3  x_4 x_6+ x_2+ x_2 x_5\nonumber\\&&+ x_2  x_4 x_6+x_1+x_1 x_6+x_1  x_3+x_1  x_2\;.
\end{eqnarray}

Since the integral Eq.~(\ref{eq:1lbox}) scales uniformly if $s_{12}$, $s_{14}$, $m_H^2$ and $m_t^2$ are all scaled by the same factor we can set one variable to be at a fixed scale and define the other in terms of their ratio to that variable. We choose $m_t^2\equiv 1$. We perform our comparison in an Euclidean region of phase-space
\begin{equation}
    -30 \leq s_{12}/m_t^2\leq -3 \,\qquad
    -30 \leq s_{14}/m_t^2\leq -3\,\qquad
    -30 \leq m_{H}^2/m_t^2\leq -3\,\qquad
\end{equation}
where $F > 0$.
\subsection{Accuracy}
To quantify the accuracy of our estimate $e$ of the integral compared with the true value $t$, we use two quantities. The first is 
\begin{equation}
    p=\log_{10}\left|\frac{e-t}{t}\right|
\end{equation}
which relates to the effective number of digits the estimates gets right. The more negative the value of $p$ is, the better the approximation. We also use
\begin{equation}
    r=\log_{10}\frac{e}{t}
\end{equation}
The best estimates have the narrowest peaks in this variable and are centered around $r=0$.

The network we use as an approximation of the primitive has the parameters listed in Table~\ref{tab:params}.
\begin{table}
\begin{centering}
    
\begin{tabular}{c||c|c|c|c|c|c|c|}
    & activation &   layers &  nodes  & parameter updates & lattice grid size & $n_{\rm repeat}$ & gain \\ \hline \hline
    $I_1$ & tanh & 4  &  100 & 160k & 100k & 5 & 1.5\\
    $I_1$ & sigmoid & 4  &  100 & 160k & 100k & 5 &2\\
    $I_2$ & tanh  & 4 & 30 & 40k & 20k & 5 & 1.2\\
    $I_2$ & sigmoid & 4 &  30 &  40k & 20k & 5 & 2\\\hline
\end{tabular}
\end{centering}
\caption{Network parameter for the results shown in this section.}\label{tab:params}
\end{table}

For the training we used a rank-one lattice and shifted it by random amounts every $n_{\rm repeat}$ network parameter updates. Keeping to a small value of $n_{repeat}$ prevents the network from overfitting to specific aspects of the chosen training sample. Since we keep generating new data there is no concept of an "epoch" in our training strategy.

The network weights are initialised using Xavier initialisation~\cite{pmlr-v9-glorot10a}, but with a gain factor dependent on the activation function, as listed in Table~\ref{tab:params}. We found that this initialisation has an important impact on the convergence of the network.

We compare the result for our integration method with numerical targets  for integral~\ref{eq:1lbox} calculated using {\tt PySecDec}~\cite{Borowka:2019zhf}. 

\begin{figure}
    \centering
    \includegraphics[scale=0.72]{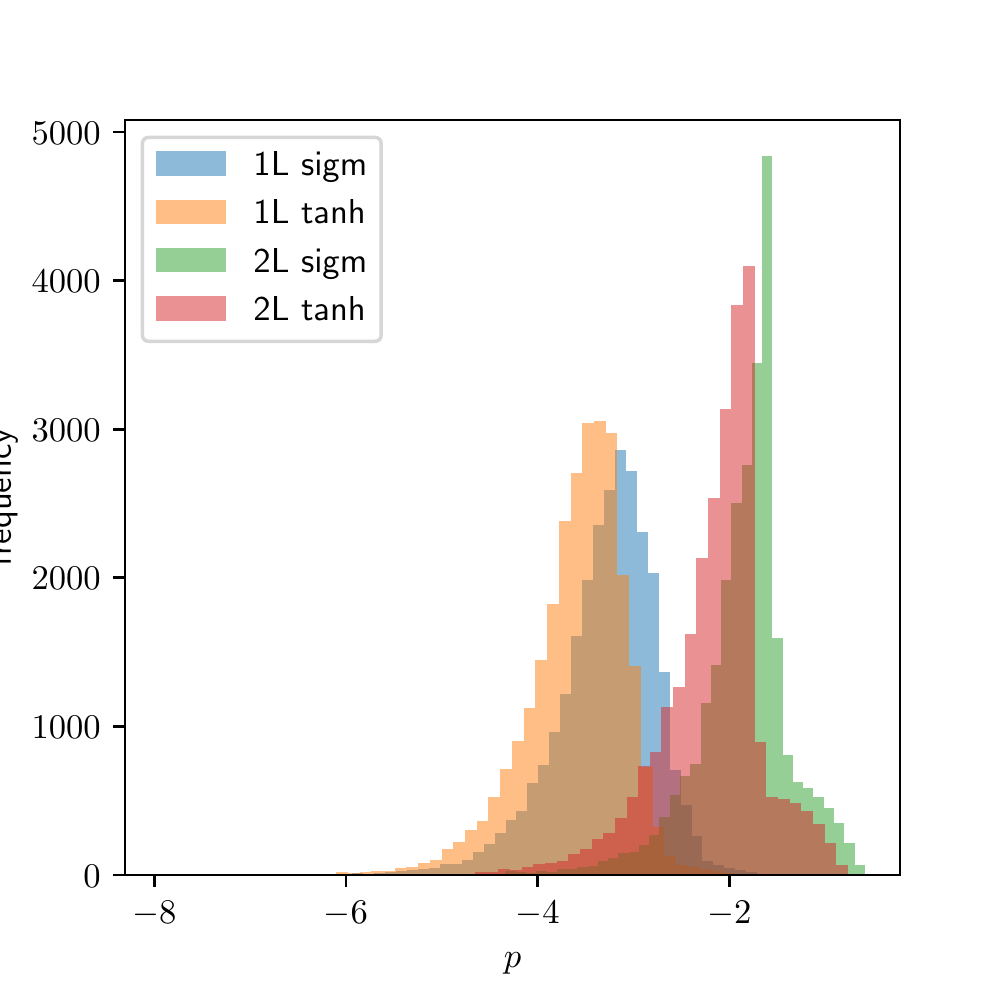}
    \includegraphics[scale=0.72]{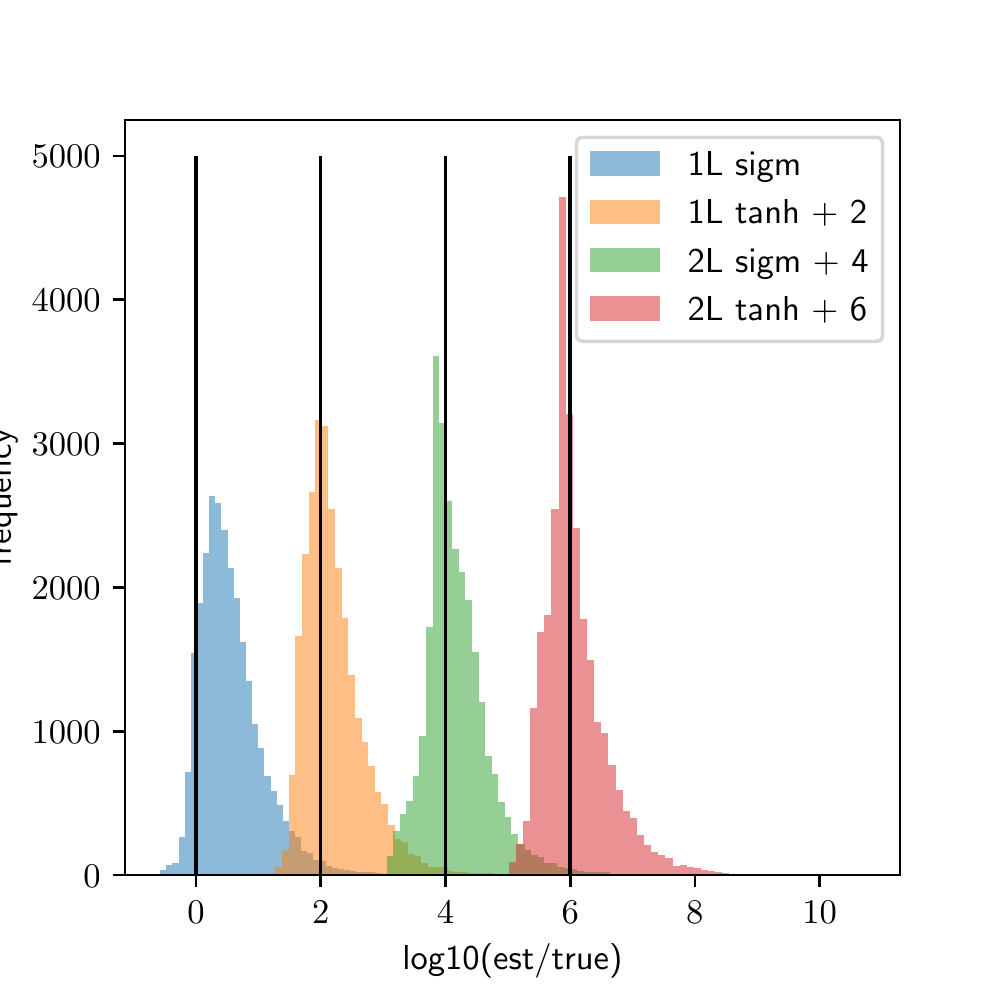}
    \caption{Left: Number of digit accuracy for the two integrals for two different activation functions. Right: Ratio of the actual error of the standard deviation of the replica results. The histograms are shifted for readability. The vertical lines correspond to the actual error matching the standard deviation of the network predictions.}
    \label{fig:errorvsstd}
\end{figure}

We trained 8 replicas of the network on the derivative data. We use the average of the individual network estimates as our prediction, and the standard deviation of the replica estimates as an uncertainty on the prediction. The accuracy is shown on the left panel of Figure~\ref{fig:errorvsstd}. The results for the two different activation functions are similar. The two-loop integral has a lower accuracy than the one-loop integral, which is to be expected given the larger number of integrations to perform (six and three, respectively). The replicas have different initialisation of their network and are trained on different samples of the integrand so the spread of their prediction encompasses both the uncertainty due to the neural network fitting and the variation in the training samples used. 

The right-hand panel of Figure~\ref{fig:errorvsstd} shows the ratio of the actual error and the estimate of the uncertainty $v$ based on the standard deviation of the replica estimates. We see that the distribution peaks around one and a small fraction of points have an error significantly higher than the estimate, showing that the errors are dominated by the variance across replica rather than a common bias. A larger number of replica would improve the accuracy of the uncertainty determination. 

\begin{figure}
    \centering
    \includegraphics[scale=0.72]{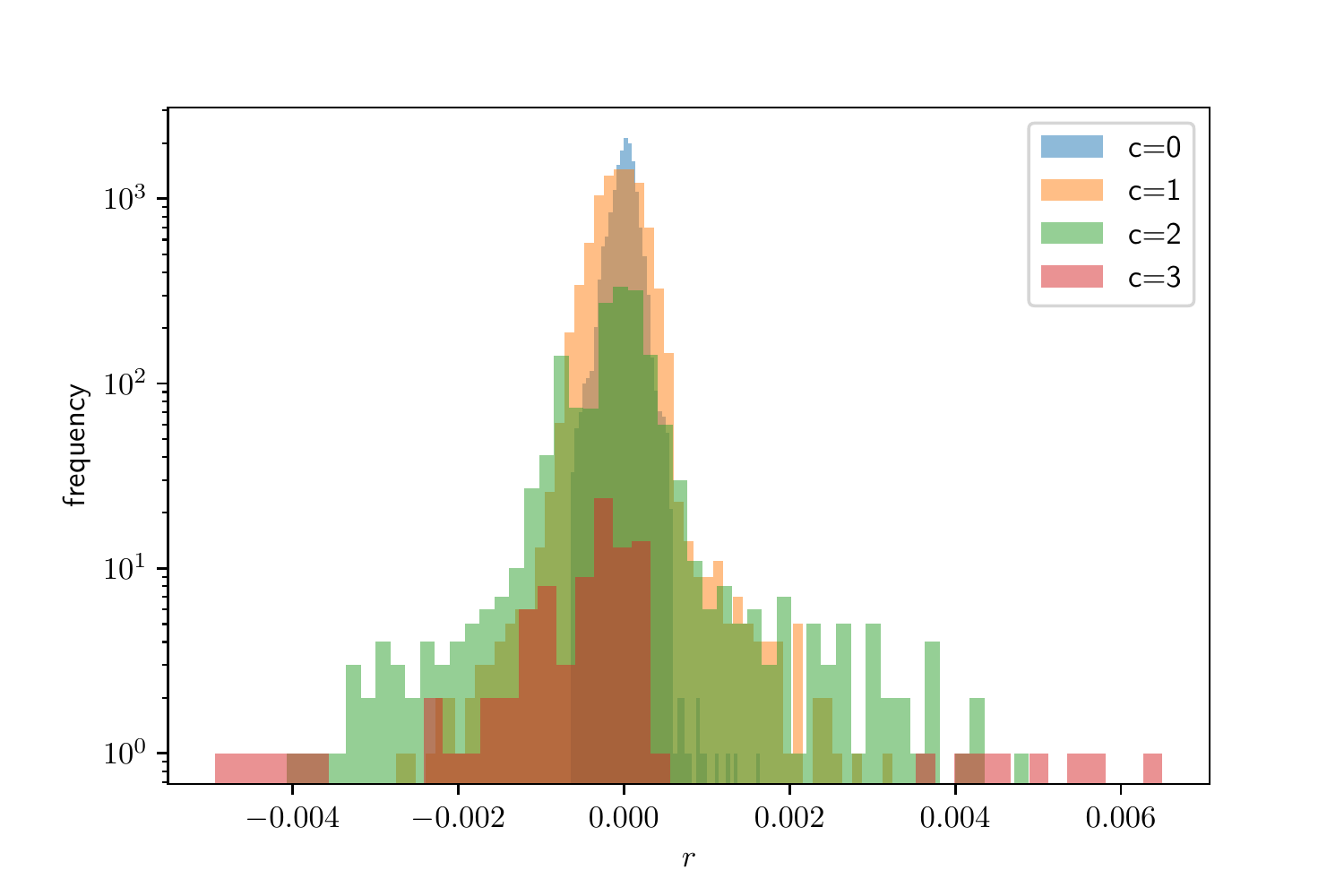}
    \caption{Distribution of the logarithm of the ratio between the estimate and the true value of the integral for different values of cornerness. The data is for integral $I_1$ and the $tanh$ activation function.}
    \label{fig:cornerness}
\end{figure}

The accuracy of the estimation is typically worse on the edges of the region of the parameters. To illustrate this we define "cornerness" $c$ as the the number of coordinates $s_{12}$, $s_{14}$ and $m_H^2$ that fall outside of the interval $[-28, -4]$. Figure~\ref{fig:cornerness} shows that the accuracy of the estimation gets worse (the distribution is wider) when one departs from the bulk of the training region.

\subsection{Timing}

Once fitted, that neural network model can provide values for the integral extremely quickly, and if evaluated on a GPU a large number of values can be evaluated in parallel. Table~\ref{tab:timings} shows approximate timings for the evaluation of the integral to a precision approximately matching that of our method.   

\begin{table}[]
    \centering
    \begin{tabular}{c|c|c|}
         & $I_1\;  \mbox{($\simeq3$ digits accuracy) }$& $I_2\;  \mbox{($\simeq2$ digits accuracy)} $\\\hline
    PySecDec & 0.8{\rm ms} &  50 {\rm ms}\\
    NN  & 0.02{\rm ms}   & 0.4 {\rm ms} 
    \end{tabular}
    \caption{Comparison for the approximate evaluation time of the integral using our method and {\tt pySecDec}.}
    \label{tab:timings}
\end{table}
The difference in timing between the one-loop and two-loop evaluation is due to the difference in the number of terms in Eq.~\ref{eq:boundarysum}. 

Evidently the time needed to train the networks has to be taken into account and amortized over the calculation of the integrals for different parameters. Currently it takes around 10 hours to train one replica for one of the integrals, so the cost of training the network starts paying off when considering the evaluation of around 360 million points at one loop and 6 million points at two loop. We expect that the performance improvements described in Section~\ref{sec:improvements} will make our method advantageous for a lower number of points.
\subsection{Possible improvements}\label{sec:improvements}
The results presented in this article are intended to demonstrate a proof of concept for the new technique for parametric integration we propose. We are convinced that the performance of the method can be greatly improved and in this section we outline a few directions to be investigated to fully exploit the potential of the method.

The results could be improved by increasing the training time, increasing the network size or increasing the number of replicas used. Increasing the number of replica improve the accuracy provided the discrepancy between the individual replica predictions has variance larger than a potential bias. Another advantage of a larger number of replica is the improvement of the error estimate, again provided the variance dominates the individual replica outputs. 

The best way to present the integrand information in the training phase should also be investigated. The training data can be drawn randomly from the $x-s$ space or can be generated from a more regular lattice. An approach similar to that used in PySecDec with a rank-one lattice shifted randomly offers more parameter for optimisation: the size of the lattice and the number of random shifts.

The optimal choice of hyperparameter for the network training are likely to be dependent on the number of auxiliary parameters to be integrated, due to the fact that as the network is differentiated more, its output becomes more oscillatory.

The initialisation of the network is also important. We found that the impact on convergence from the initialisation weights is more pronounced with our loss function than in a regular MSE loss due to the fact that the values of the derivatives are larger than those of the original network, leading to more risk of exponential rise or dampening of the node activations as we propagate through the network.  

We leave the detailed optimisation of the precision trade-offs between network capacity and training time to a future study.

\section{Conclusion}\label{sec:conclusion}

In this article we presented a method to perform multi-variable integration using a neural-network fit of the primitive of the integrand. Having an approximation for the primitive of an integrand function $f$ can have additional advantages beyond obtaining the integral over all auxiliary parameters. If $f$ is a probability density, one could use $\mathcal{N}$ to calculate marginal distributions or conditional probabilities $n(s_1,...s_m; x_i| x_1,.,x_{i-1}, x_{i+1},...,x_k)$ and draw unweighted samples using Gibbs sampling~\cite{Gibbs}. 

The method presented in this article is particularly suited to cases where not all arguments of the integrand function are integrated over. We showed that without much efforts put in optimising the network and its training we can obtain a reasonable accuracy for the integral.   
There is much to explore about the training of a network with a loss function involving its derivatives to multiple degrees. We are convinced that both the outcomes of the method presented in this article and the efficiency they are arrived at can be greatly improved with a more systematic study. A better understanding will be key to embark on more complicated application, including higher number of auxiliary parameters and evaluations in Minkowsky space.

\section*{Acknowledgments}
We would like to thank Stephen Jones for useful discussions on the Korobov transform, quasi-Monte Carlo grids and drafts of this article, and also for providing us with the target values for the examples. We are also thankful to Carina Popovici for insightful discussion on the convergence of our network with its particular loss function. DM would like to express special thanks to the Mainz Institute for Theoretical Physics (MITP) of the Cluster of Excellence PRISMA* (Project ID 39083149) for its hospitality and support. RS-M would like to thank Art Recognition AG for their hospitality and for allowing us to use their cloud infrastructure for some initial experiments.
\bibliography{notes}
\bibliographystyle{JHEP}

\appendix

\section{Derivatives of the sigmoid and $tanh$ function}\label{sec:sigmoidderivatives}

Here we list the derivatives of the sigmoid function $f(x)=\frac{1}{1+e^-x}\equiv y$:

\begin{eqnarray}
    f^{(1)}(x)&=&(1 - y) y \nonumber\\
    f^{(2)}(x)&=& (1 - y) y (1 - 2 y)\nonumber\\
    f^{(3)}(x)&=& (1 - y) y (1 - 6 y + 6 y^2)\nonumber\\
    f^{(4)}(x)&=& (1 - y) y (1 - 2 y) (1 - 12 y + 12 y^2)\nonumber\\
    f^{(5)}(x)&=& (1 - y) y (1 - 30 y + 150 y^2 - 240 y^3 + 120 y^4)\nonumber\\
    f^{(6)}(x)&=& (1 - y) y (1 - 2 y) (1 - 60 y + 420 y^2 - 720 y^3 + 
    360 y^4)\nonumber\\
    f^{(7)}(x)&=& (1 - y) y (1 - 126 y + 1806 y^2 - 8400 y^3 + 
      16800 y^4 - 15120 y^5 + 5040 y^6)\nonumber\\
    f^{(8)}(x)&=& (1 - y) y (1 - 2 y) (1 - 
    252 y + 5292 y^2 - 30240 y^3 + 65520 y^4 - 60480 y^5 + 
    20160 y^6)\nonumber\\
    f^{(9)}(x)&=& (1 - y) y (1 - 510 y + 18150 y^2 - 186480 y^3 + 
      834120 y^4 - 1905120 y^5 + 2328480 y^6 \nonumber\\ &&- 1451520 y^7 + 
      362880 y^8)\nonumber\\
      f^{(10)}(x) &=& (1 - y) y (1 - 2 y) (1 - 1020 y + 53940 y^2 - 
    710640 y^3 + 3681720 y^4 - 9072000 y^5 \nonumber\\&&+ 11491200 y^6 - 
    7257600 y^7 + 1814400 y^8)\nonumber
\end{eqnarray}

The derivatives of the $tanh$ functions are listed here:

$f(x)=\tanh(x)\equiv t$
\begin{eqnarray}
   f^{(1)}(x)&=&(1 - t) (1 + t))\nonumber\\ 
   f^{(2)}(x)&=& - 2 (1 - t) t (1 + t)\nonumber\\ 
   f^{(3)}(x)&=& - 2 (1 - t) (1 + t) (1 - 3 t^2)\nonumber\\
   f^{(4)}(x)&=&8 (1 - t) t (1 + t) (2 - 3 t^2)\nonumber\\
   f^{(5)}(x)&=& 8 (1 - t) (1 + t) (2 - 15 t^2 + 
    15 t^4)\nonumber\\
   f^{(6)}(x)&=&
 -16 (1 - t) t (1 + t) (17 - 60 t^2 + 45 t^4)\nonumber\\
   f^{(7)}(x)&=& 16 (1 - t) (1 + 
    t) (-17 + 231 t^2 - 525 t^4 + 315 t^6)\nonumber\\ 
   f^{(8)}(x)&=& -128 (1 - t) t (1 + t) (-62 + 378 t^2 - 630 t^4 + 
    315 t^6)\nonumber\\ 
    f^{(9)}(x)&=& 128 (1 - t) (1 + t) (62 - 1320 t^2 + 5040 t^4 - 
    6615 t^6 + 2835 t^8)\nonumber\\ 
    f^{(10)}(x)&=& -256 (1 - t) t (1 + t) (1382 - 12720 t^2 + 34965 t^4 - 37800 t^6 + 
    14175 t^8)\nonumber 
\end{eqnarray}

\section{Further derivatives}\label{sec:derivatives}
For a parametric integral with $k$ integrated variables we need to differential the network $k$ times in the loss function. The expressions are obtained through repeated application of the chain rule. In this section we explicitly show the expression for the third and fourth derivative of the network output with respect to its input. 

The derivatives of the activation functions are easy to generalise:
\begin{equation}
    \frac{d^pz^{(l)}_i}{dx_1dx_2...dx_p} = \sum\limits_j w_{ij}^{(l)}\frac{d^pa^{(l-1)}_i}{dx_1dx_2...dx_p}
\end{equation}
and
\begin{equation}
    \frac{d^pa^{(0)}_i}{dx_1dx_2...dx_p} = 0\;,\qquad p > 1
\end{equation}
The derivatives of the activation values is less easy to express in general, we give here the explicit expression for three and four derivatives. The third derivative with respect to $x_3$ is given by
\begin{eqnarray}
    \frac{d^3a^{(l)}_i}{dx_1dx_2dx_3} &=& \phi^{(3)}(z^{(l)}_i) \frac{dz_i^{(l)}}{dx_1}\frac{dz_i^{(l)}}{dx_2}\frac{dz_i^{(l)}}{dx_3} \nonumber\\
    &&+\phi''(z^{(l)}_i)\left[ \frac{d^2z_i^{(l)}}{dx_1dx_3}\frac{dz_i^{(l)}}{dx_2} +
     \frac{dz_i^{(l)}}{dx_1}\frac{d^2z_i^{(l)}}{dx_2dx_3} +
     \frac{d^2z_i^{(l)}}{dx_1dx_2}\frac{dz_i^{(l)}}{dx_3}\right] \nonumber\\
    &&+\phi'(z^{(l)}_i) \frac{d^3z_i^{(l)}}{dx_1dx_2dx_3} \nonumber\\
\end{eqnarray} 
The fourth derivative with respect to $x_4$ is given below.
\begin{eqnarray}
    \lefteqn{\frac{d^4a^{(l)}_i}{dx_1dx_2dx_3dx_4} =}&&\nonumber\\&& \phi^{(4)}(z^{(l)}_i) \frac{dz_i^{(l)}}{dx_1}\frac{dz_i^{(l)}}{dx_2}\frac{dz_i^{(l)}}{dx_3}\frac{dz_i^{(l)}}{dx_4} \nonumber\\
    &&+\phi^{(3)}(z^{(l)}_i)\left[ 
     \frac{d^2z_i^{(l)}}{dx_1dx_2}\frac{dz_i^{(l)}}{dx_3}\frac{dz_i^{(l)}}{dx_4} +
     \frac{d^2z_i^{(l)}}{dx_1dx_3}\frac{dz_i^{(l)}}{dx_2}\frac{dz_i^{(l)}}{dx_4} 
     +\frac{d^2z_i^{(l)}}{dx_1dx_4}\frac{dz_i^{(l)}}{dx_2}\frac{dz_i^{(l)}}{dx_3}\right. \nonumber\\
     &&\left.+\frac{d^2z_i^{(l)}}{dx_2dx_3}\frac{dz_i^{(l)}}{dx_1}\frac{dz_i^{(l)}}{dx_4} 
     +\frac{d^2z_i^{(l)}}{dx_2dx_4}\frac{dz_i^{(l)}}{dx_1}\frac{dz_i^{(l)}}{dx_3} +
     \frac{d^2z_i^{(l)}}{dx_3dx_4}\frac{dz_i^{(l)}}{dx_1}\frac{dz_i^{(l)}}{dx_2} 
     \right] \nonumber\\
     &&+\phi^{(2)}\left[
        \frac{d^2 z_i^{(l)}}{dx_1 dx_2}\frac{d^2 z_i^{(l)}}{dx_3 dx_4} 
        +\frac{d^2 z_i^{(l)}}{dx_1 dx_3}\frac{d^2 z_i^{(l)}}{dx_2 dx_4} 
        +\frac{d^2 z_i^{(l)}}{dx_1 dx_4}\frac{d^2 z_i^{(l)}}{dx_2 dx_3} \right] \nonumber\\
    &&+\phi^{(2)}\left[
        \frac{d^3 z_i^{(l)}}{dx_1 dx_2 dx_3}\frac{d z_i^{(l)}}{dx_4} 
        +\frac{d^3 z_i^{(l)}}{dx_1 dx_2 dx_4}\frac{d z_i^{(l)}}{dx_3} 
        +\frac{d^3 z_i^{(l)}}{dx_1 dx_3 dx_4}\frac{d z_i^{(l)}}{dx_2} 
        +\frac{d^3 z_i^{(l)}}{dx_2 dx_3 dx_4}\frac{d z_i^{(l)}}{dx_1} 
         \right] \nonumber\\
        &&+\phi'(z^{(l)}_i) \frac{d^4z_i^{(l)}}{dx_1dx_2dx_3dx_4} \nonumber\\
\end{eqnarray}

\end{document}